\documentclass{iau}
\usepackage{graphicx}

\title[Evolution of BCGs] %% give here short title %%
{Updated catalog of 132,684 galaxy clusters and evolution of brightest cluster galaxies}

\author[Z. L. Wen \& J. L. Han]   %% give here short author list %%
{Z. L. Wen
 \and J. L. Han}

\affiliation{National Astronomical Observatories, 
                Chinese Academy of Sciences, 
                20A Datun Road, Chaoyang District, Beijing 100012, China; 
                zhonglue@bao.ac.cn.}

\pubyear{2013}
\volume{295}  %% insert here IAU Symposium No.
\pagerange{xx--xx}
\jname{The intriguing life of massive galaxies}
\editors{D. Thomas, A. Pasquali \& I. Ferreras, eds.}
\begin{document}

\maketitle

\begin{abstract}
We identified 132,684 clusters in the redshift range of $0.05<z<0.8$
from SDSS DR8.  The spectroscopic redshifts of 52,683 clusters have
been included in the catalog using SDSS DR9 data. We found that BCGs
are more luminous in richer clusters and at higher redshifts.
%\keywords{clusters of galaxies.}
\end{abstract}

%\section{Introduction}

From the Sloan Digital Sky Survey Data Release 8 (SDSS DR8), we
identified 132,684 clusters in the redshift range of $0.05<z<0.8$
(\cite[Wen et al. 2012]{whl12}). Using photometric redshifts of
galaxies, we recognized a cluster when the richness $R_{L*}=L_{\rm
  total}/L^{\ast}\ge12$ and the number of member galaxy candidates
within a photometric redshift gap of $z\pm0.04(1+z)$ and a radius of
$r_{200}$, $N_{200}\ge8$. Here, $r_{200}$ is the radius within which
the mean density of a cluster is 200 times of the critical density of
the universe, $L_{\rm total}$ is the total luminosity of member
galaxies, $L^{\ast}$ is the characteristic luminosity. Monte Carlo
simulations show that for rich clusters with a mass of
$M_{200}>1.0\times10^{14}~M_{\odot}$ in the redshift range of $0.05\le
z< 0.42$, the sample has a purity of $>95\%$ and a completeness of
$>95\%$. The spectroscopic redshifts of 52,683 clusters have been
included in the catalog based on the spectra data of the SDSS DR9. The
updated catalog of galaxy clusters is now available at
http://zmtt.bao.ac.cn/galaxy$\_$clusters/.

Brightest cluster galaxies (BCGs) are luminous elliptical galaxies
located at the potential centers of galaxy clusters.  Because of the
dominant role inside clusters and their unusual properties, the
formation and evolution of BCGs are very intriguing.  The BCGs of the
clusters in our catalog are recognized as the brightest member
galaxies within a radius of 0.5 Mpc from the number density peaks. We
found that BCGs are more luminous in richer clusters and at higher
redshifts, in the form of
\\ $M_r=(-21.25\pm0.01)-(1.75\pm0.03)z-(1.10\pm0.03)\log
R_{L\ast}$.\\ The color evolution of BCGs was investigated using a
higher redshift cluster sample identified from the
Canada-France-Hawaii Telescope Deep Survey and the Cosmic Evolution
Survey using photometric redshifts (\cite[Wen \& Han
  2011]{wh11}). There are 294 clusters in the redshift range of
$0.5<z<1.6$. The colors $r'-z'$ and $r^+-m_{3.6\mu m}$ of most BCGs
are consistent with a stellar population synthesis model
(\cite[Bruzual \& Charlot 2003]{bc03}) in which the BCGs are formed at
redshift $z_f>2$ and are evolved passively.

We also checked if the BCGs are luminous red galaxies (LRGs). We found
that 25\% of LRGs are the BCGs of our clusters, and 36\% of LRGs are
cluster member galaxies. In our cluster sample, 63\% of BCGs satisfy
the SDSS LRGs selection criteria for magnitude of $r_{\rm
  petro}<19.5$.

\end{document}